\documentclass{ws-ijmpd}
\usepackage[super,compress]{cite}
\usepackage[unicode]{hyperref}
\begin{document}
\markboth{T. Q. Do and S. H. Q. Nguyen}
{Anisotropic power-law inflation in a two-scalar-field model with a mixed kinetic term}

%
\catchline{}{}{}{}{}
%

\title{Anisotropic power-law inflation in a two-scalar-field model with a mixed kinetic term}

\author{Tuan Q. Do}

\address{Faculty of Physics, VNU University of Science, Vietnam National University, \\
Hanoi 120000, Vietnam\\
tuanqdo@vnu.edu.vn}

\author{Sonnet Hung Q. Nguyen}

\address{Faculty of Physics, VNU University of Science, Vietnam National University, \\
Hanoi 120000, Vietnam\\
hungnq\_kvl@vnu.edu.vn}

\maketitle

\begin{history}
\received{Day Month Year}
\revised{Day Month Year}
\end{history}

\begin{abstract}
We examine whether an extended scenario of a two-scalar-field model, in which a mixed kinetic term of canonical and phantom scalar fields is involved, admits the Bianchi type I metric, which is homogeneous but anisotropic spacetime, as its power-law solutions. Then we analyze the stability of the  anisotropic power-law solutions to see whether these solutions respect the cosmic no-hair conjecture or not during the inflationary phase. In addition, we will also investigate a special scenario, where the pure kinetic terms of canonical and phantom fields disappear altogether in field equations, to test again the validity of  cosmic no-hair conjecture. As a result, the cosmic no-hair conjecture always holds in both these scenarios due to the instability of the corresponding anisotropic inflationary solutions.
\end{abstract}


\ccode{PACS numbers: 98.80.-k; 98.80.Cq; 98.80.Jk}
\section{Introduction}\label{sec1}
An inflationary universe \cite{Guth} has been considered as a leading paradigm in modern cosmology due to its power of solving some classical cosmological problems such as the horizon, flatness, and magnetic-monopole problems as well as its consistent predictions for the cosmic microwave background.  In particular, many theoretical predictions based on the inflationary mechanism have been shown to be highly consistent with recent high-tech observations on the cosmic microwave background (CMB) such as the Wilkinson Microwave Anisotropy Probe (WMAP) \cite{WMAP} or Planck \cite{Planck}. However, some exotic features of the CMB temperature like the hemispherical asymmetry and the Cold Spot have been firstly observed by the WMAP \cite{WMAP} and then confirmed by the Planck \cite{Planck}. Hence, the nature of these anomalies requires further investigations, which might address some additional or unusual interactions of fields, e.g., those might come from string theories \cite{Chernoff:2014cba}. Due to these anomalies, our imagination of the early universe, which has been thought of  being homogeneous and isotropic, might be changed  \cite{Buchert:2015wwr}. For example, we might think of a scenario that the state of the  early universe might be described by the Bianchi spacetimes rather than the Friedmann-Lemaitre-Robertson-Walker (FLRW) one since it might be not isotropic but slightly anisotropic \cite{Ade:2013vbw}. In cosmology, the Bianchi spacetimes are known as homogeneous but anisotropic metrics, which are classified into nine types numbered from I to IX \cite{Bianchi} and are regarded as the generalization of the FLRW metric, which is homogeneous and isotropic. Recently, cosmological aspects based on the Bianchi spacetimes have been studied extensively. For example, some early works on the predictions of  an anisotropic inflationary era can be found in Ref. \refcite{Pitrou:2008gk}. In addition, other works within the framework of loop quantum cosmology (gravity) on understanding the issues of the resolution of initial singularity, effect of anisotropies on inflation,  isotropization, and stablitily of inflationary attractors for the Bianchi type I metric have been investigated in Ref. \refcite{Singh:2011gp}.

As mentioned above, the data of WMAP and Planck  might have shown us the state of the early universe, which might be anisotropic with small spatial anisotropies. Naturally, we can question on the state of the late time universe. "Is it isotropic or not" is an open question to all of us, which could be answered by theoretical and/or observational approaches. Fortunately, an important theoretical hint to this question might come from the cosmic no-hair conjecture proposed by Hawking and his colleagues long time ago \cite{Hawking}, which states that all classical hairs of the early universe will be removed at the late time. It is noted that a complete proof for this conjecture has not been done up to now. However, a partial proof dealing with the dominant energy condition (DEC) and strong energy condition (SEC) for all non-type-IX  Bianchi spacetimes has been given by  Wald in Ref. \refcite{wald}. As a result, this proof shows that  all non-type-IX Bianchi spacetimes will evolve towards the late time isotropic de Sitter spacetime if  the DEC and SEC are both fulfilled. For the Bianchi type IX metric, it will behave similarly if the cosmological constant $\Lambda$ is sufficiently large \cite{wald}. Recently, some people have tried to extend the Wald's proof to  a case of inhomogeneous cosmologies \cite{inhomogeneous}. Indeed, a complete proof for this conjecture has been one of great challenges to physicists. In short, if the cosmic no-hair conjecture holds, the late time state of universe should be isotropic, no matter the early state of universe. 

However, the cosmic no-hair conjecture has faced counter-examples coming from a supergravity motivated model proposed by Kanno, Soda, and Watanabe (KSW)\cite{KSW,KSW1}, where  a unusual coupling of the scalar $\phi$ and $U(1)$ fields, $f^2(\phi)F_{\mu\nu}F^{\mu\nu}$, is involved. As a result, the KSW model does admit  Bianchi type I metrics as its stable and attractor solutions during the inflationary phase. More interestingly, this result still holds when a canonical scalar field $\phi$ is replaced by non-canonical ones, e.g., the (supersymmetric-) Dirac-Born-Infeld scalar fields, as shown in Ref. \refcite{WFK}. Hence, the cosmic no-hair conjecture seems to be  violated extensively in the context of the KSW model. Consequently, there have been a number of papers investigating possible extensions of the KSW model to seek more counter-examples to the cosmic no-hair conjecture  \cite{extensions}. Additionally, some cosmological aspects such as imprints of anisotropic inflation on the CMB through correlations between $T$, $E$, and $B$ modes\cite{implication}, and primordial gravitational waves\cite{ito}  have also been discussed in the framework of KSW model. For recent interesting reviews on this model, see Ref. \refcite{KSW1}.

Besides the above counter-examples, there have existed some papers \cite{WFK} attempting to support the cosmic no-hair conjecture by introducing a unusual scalar field called a phantom field $\psi$, whose kinetic energy is negative definite \cite{phantom,Singh:2003vx,phantom1,quintom,quintom1,quintom2,quintom3}. Cosmologically, the phantom field has been regarded as one of alternative solutions to the dark energy problem, which is associated with the accelerating of our current universe \cite{phantom,Singh:2003vx,quintom,quintom1,Copeland:2006wr}. For example, one can see the very first  confirmations for the cosmological viability of phantom field model done in Ref. \refcite{Singh:2003vx}.  However, the existence of phantom field has been shown to lead the Universe dominated by the phantom energy to the so-called Big Rip singularity, which is the finite-time future singularity\cite{phantom1}.  Fortunately, a two-scalar-field model called a quintom model, which includes not only the phantom field but also the quintessence field, has provided alternative solutions to not only the Big Rip singularity but also other cosmological singularities, which have been discussed extensively in loop quantum cosmology \cite{Singh:2011gp,singularity,Ashtekar:2011ni}, such as the Big Bang  singularity \cite{quintom2} and  the Big  Crunch   singularity\cite{quintom3}.  On the other hand, papers in Ref. \refcite{Singh} have pointed out that the Big Rip singularity problem associated with the existence of phantom field can be resolved once the quantum gravitational effects are involved. 

These facts make the role of phantom field in cosmology  important. Indeed, the phantom has been shown that its existence might also be necessary  for supporting the cosmic no-hair conjecture \cite{WFK}. In particular, the stability analysis done in  Ref. \refcite{WFK} has shown that the inclusion of phantom field does make the following anisotropic Bianchi type I solutions unstable during the inflationary phase due to the negativity of its kinetic energy as expected. However, one could ask if this result would still be valid if additional unusual terms of scalar fields are introduced into the two-scalar-field model \cite{WFK}. In the present paper, we will partially answer this question by examining an extended scenario of the two-scalar-field model \cite{WFK}, in which a mixed kinetic term of canonical and phantom scalar fields, i.e., $\partial_\mu \phi \partial^\mu \psi$ \cite{mixed}, is involved. Note that this mixed kinetic term can also be found in the  string motivated models of multi scalar fields \cite{mixed-1}. As a result, this mixed term will not make the corresponding Bianchi type I solutions stable during the inflationary phase. Furthermore, we will  show that this result is also valid for a special scenario, in which the pure (non-mixed) kinetic terms of canonical and phantom fields are neglected altogether. Indeed, it will be shown that the corresponding anisotropic power-law solutions found in this scenario also turn out to be unstable as expected. 

This paper is organized as follows:  A brief introduction of this research has been given in Sec. \ref{sec1}. A two-scalar-field model with the mixed term and its anisotropic power-law solutions will be solved in Sec. \ref{sec2}. Stability of the anisotropic power-law solutions will be analyzed in Sec. \ref{sec3} to see whether the cosmic no-hair conjecture is violated or not.  Sec. \ref{sec4} will be devoted to investigate a special scenario, in which the pure kinetic terms of canonical and phantom fields are neglected altogether.  Finally, concluding remarks will be given in Sec. \ref{sec5}.
\section{The model and its anisotropic power-law solutions} \label{sec2}
\subsection{Basic setup}
An action of an extended scenario of KSW model \cite{KSW} including the phantom field \cite{WFK,phantom,Singh:2003vx,phantom1,quintom,quintom1} and the mixed kinetic term \cite{mixed,mixed-1} is given by
\begin{align} \label{action}
  S = \int {d^4 } x\sqrt { - g} & \left[ {\frac{M_p^2}
{2}R -a \partial ^\mu  \phi \partial _\mu  \phi + b \partial ^\mu  \psi \partial _\mu  \psi -\frac{\omega_0}{2} \partial^\mu \phi \partial_\mu \psi }\right. \nonumber\\
 & ~\left.{- V_\phi(\phi) -V_\psi(\psi) - \frac{1}{4}f^2( \phi,\psi )F_{\mu \nu } F^{\mu \nu } } \right],
\end{align}
where $a\geq0$ and $b\geq0$ are coefficients of the kinetic term of  canonical scalar $\phi$ and phantom scalar $\psi$ fields, respectively.  In addition, $\omega_0$ is a coefficient of mixed kinetic term. It appears that for $\omega_0>0$ and $\omega_0 <0$ we will have the quintessence-like and phantom-like mixed terms, respectively. In addition, $M_p$ is the reduced Planck mass and $F_{\mu\nu}\equiv \partial_\mu A_\nu -\partial_\nu A_\mu$ is the field strength of the vector field $A_\mu$ used for describing the electromagnetic field. In addition, it appears that $F^{\mu\nu}=g^{\mu\rho}g^{\nu\sigma}F_{\rho\sigma}$. Note that if $\omega_0=0$  we will obtain the two-scalar-field model studied in Ref. \refcite{WFK}.

As a result, varying the action (\ref{action}) with respect to the inverse metric $g^{\mu\nu}$ and choosing the canonical coefficients, $a=b=1/2$, lead to the following Einstein field equations:
\begin{align} \label{Einstein-equations}
 & M_p^2\left( {R_{\mu \nu }  - \frac{1}
{2}Rg_{\mu \nu } } \right) - \partial _\mu  \phi \partial _\nu  \phi  + \partial _\mu  \psi \partial _\nu  \psi -\omega_0 \partial_\mu \phi \partial_\nu \psi \nonumber\\
&+ \frac{1}{2}g_{\mu \nu }\left( \partial ^\sigma  \phi \partial _\sigma  \phi -  \partial ^\sigma  \psi \partial _\sigma  \psi +\omega_0 \partial ^\sigma  \phi \partial _\sigma  \psi \right) \nonumber\\
&   + g_{\mu \nu } \left[ {V_\phi\left( \phi \right) + V_\psi\left(\psi \right)+ \frac{1}
{4}f^2 \left( \phi,\psi  \right)F^{\rho \sigma } F_{\rho \sigma } } \right] - f^2 \left( \phi ,\psi \right)F_{\mu \gamma } F_\nu ^\gamma   = 0,
\end{align}

Additionally, the Euler-Lagrange equations for the scalar fields, $\phi$ and $\psi$, and the vector field $A_\mu$ read
\begin{eqnarray} \label{phi-equation}
  \ddot \phi +\frac{\omega_0}{2}\ddot\psi  &=&  - 3H \left(\dot \phi +\frac{\omega_0}{2} \dot\psi \right)  - \partial_\phi V_\phi \left( \phi  \right) - \frac{1}{2}f\left( \phi ,\psi \right) \partial_\phi f \left( \phi ,\psi \right)F_{\mu \nu } F^{\mu \nu } , \\
\label{psi-equation}
  \ddot \psi -\frac{\omega_0}{2}\ddot\phi   &=&  - 3H \left(\dot \psi -\frac{\omega_0}{2} \dot\phi \right)  + \partial_\psi V_\psi \left(\psi  \right) + \frac{1}{2}f\left( \phi,\psi  \right) \partial_\psi f \left( \phi ,\psi \right)F_{\mu \nu } F^{\mu \nu } , 
\end{eqnarray}
\begin{equation} \label{vector-equation}
\frac{\partial }
{{\partial x^\mu  }}\left[ {\sqrt { - g} f^2 \left( \phi,\psi  \right)F^{\mu \nu } } \right] = 0,
\end{equation}
respectively, where $H$ is the Hubble constant appearing due to the derivative of $\sqrt{-g}$, i.e., $\partial_\mu\left(\sqrt{-g}\right) $.

Given the general forms of field equations, we would like to seek analytic solutions for the two-scalar-field model with the mixed term as described in the action (\ref{action}) by following the previous works in  Refs. \refcite{KSW,WFK}. In particular, we are now interested in a question that whether the two-scalar-field model involving the mixed term  admits the Bianchi type I (BI) metric:
\begin{equation}\label{bianchi-I}
 ds^2 =  -dt^2+\exp\left[{2\alpha(t)-4\sigma(t)}\right]dx^2+\exp\left[{2\alpha(t)+2\sigma(t)}\right]\left({dy^2+dz^2}\right)  , 
\end{equation}
along with the compatible vector field, whose configuration is given by $A_\mu = (0,A_x(t),0,0)$, as its cosmological solutions. Here, $\sigma$ stands for a deviation from isotropy and therefore should be much smaller than the isotropic scale factor $\alpha$ in order to be consistent with the recent observation data from the WMAP \cite{WMAP} and Planck \cite{Planck}. 
Note that among nine Bianchi types, the Bianchi type I seems to be closest to the FLRW metric since its metric is diagonal, similar to the FLRW metric. This is a reason why the Bianchi type I metric has been investigated extensively \cite{Pitrou:2008gk,Singh:2011gp,singularity,Ashtekar:2011ni,KSW1,TQD,extensions}.  Now, we would like to derive the corresponding field equations (\ref{Einstein-equations}), (\ref{phi-equation}), (\ref{psi-equation}), and (\ref{vector-equation}) for the   BI metric shown in Eq. (\ref{bianchi-I}). To do this task, we first define the following solution for the vector field equation (\ref{vector-equation})  to be
\begin{equation}  \label{vector-equation-solution}
\dot A_x \left( t \right) = f^{ - 2} \left( \phi,\psi  \right)\exp[{ - \alpha  - 4\sigma }] p_A ,
\end{equation}
where $p_A $ is a constant of integration \cite{KSW}. Thanks to this solution, we are able to write down the non-vanishing components of Einstein equations (\ref{Einstein-equations}) as follows
\begin{align}  \label{Friedmann-equation}
   \dot \alpha ^2  &= \dot \sigma ^2  + \frac{{1 }}{3  M_p^2}\left[ {\frac{1}{2}\dot \phi ^2  -\frac{1}{2}\dot \psi ^2 +\frac{\omega_0}{2}\dot\phi \dot\psi+ V_\phi +V_\psi + \frac{f^{ - 2}}{2}\exp[{ - 4\alpha  - 4\sigma }] p_A^2 } \right] ,\\
\label{alpha-equation}
\ddot \alpha & =  - 3\dot \alpha ^2  + \frac{1 }{M_p^2} \left(V_\phi +V_\psi\right)+ \frac{f^{ - 2} }{6M_p^2} \exp[{ - 4\alpha  - 4\sigma }] p_A^2 ,\\
\label{sigma-equation}
\ddot \sigma &  =  - 3\dot \alpha \dot \sigma  + \frac{f^{ - 2}}{3 M_p^2}\exp[{ - 4\alpha  - 4\sigma }] p_A^2.
\end{align} 
On the other hand, the scalar field equations (\ref{phi-equation}) and (\ref{psi-equation}) can be reduced to
\begin{align} \label{scalar-field-equation-1}
 \ddot \phi +\frac{\omega_0}{2}\ddot\psi & =  - 3\dot\alpha \left(\dot \phi +\frac{\omega_0}{2} \dot\psi \right)  - \partial_\phi V_\phi  + f^{ - 3} \partial_\phi f \exp[{ - 4\alpha  - 4\sigma }] p_A^2 , \\
\label{scalar-field-equation-2}
\ddot \psi -\frac{\omega_0}{2}\ddot\phi   &=  - 3\dot\alpha \left(\dot \psi -\frac{\omega_0}{2} \dot\phi \right)   + \partial_\psi V_\psi  - f^{ - 3} \partial_\psi f \exp[{ - 4\alpha  - 4\sigma }] p_A^2 .
\end{align}
Note again, once we take $\omega_0 =0$ then all above equations will reduce to that investigated in the two-scalar-field model in Ref. \refcite{WFK}. As a result, the following evolution equation associated with the scale factor $\alpha$, which governs the evolution of inflationary universe since it is assumed to be much larger than the anisotropic deviation $\sigma$, turns out to be 
\begin{equation}
\ddot\alpha +\dot\alpha^2 = -2\dot\sigma^2 -\frac{1}{3M_p^2} \left(\dot\phi^2 +\omega_0 \dot\phi\dot\psi -\dot\psi^2 - V_\phi - V_\psi +\frac{f^{ - 2}}{2}\exp[{ - 4\alpha  - 4\sigma }] p_A^2  \right).
\end{equation}
For an inflationary universe it requires that $\ddot\alpha +\dot\alpha^2 >0$. This constraint will be easily fulfilled if the slow-roll approximation, in which the potentials of scalar fields dominate over other terms in the field equations, i.e., $V_\phi \gg \dot\phi^2/2$, $V_\psi \gg \dot\psi^2/2$, and $V_\phi+V_\psi \gg f^{ - 2}\exp[{ - 4\alpha  - 4\sigma }] p_A^2/{2}$, is taken. It is noted again that the anisotropic scale factor $\sigma$  must be much smaller than the isotropic scale factor $\alpha$  in order to be consistent with the recent observation data from the WMAP \cite{WMAP} and Planck \cite{Planck}.
\subsection{Anisotropic power-law solutions}
Armed with the basic setup for the Bianchi type I metric derived above, we would like to seek the power-law solutions for the two-scalar-field model with the mixed term by taking the following ansatz used in Refs. \refcite{KSW,WFK}:
\begin{equation} \label{ansatz}
\alpha = \zeta \log \left( t \right), ~ \sigma  = \eta \log \left( t \right), \frac{\phi }
{{M_p }}  = \xi_\phi \log \left( t \right) + \phi _0, ~\frac{\psi }
{{M_p }}  = \xi_\psi \log \left( t \right) + \psi _0,
\end{equation}
along with the compatible exponential  potentials: 
\begin{align} \label{expo-1}
V_\phi(\phi) &= V_{0\phi} \exp\left[{\lambda_\phi\frac{\phi }
{{M_p }} }\right] , \\
\label{expo-2}
V_\psi(\psi) &= V_{0\psi} \exp\left[{\lambda_\psi \frac{\psi}
{{M_p }} }\right] ,\\
\label{expo-3}
f\left( \phi, \psi  \right) & = f_0 \exp \left[{\rho_\phi \frac{\phi}{{M_p }} +\rho_\psi \frac{\psi  }{{M_p }}}\right],
\end{align}
where $V_{0\phi}$,  $V_{0\psi}$, $f_0$, $\lambda_\phi$, $\lambda_\psi$, $\rho_\phi$, and $\rho_\psi$ are positive field parameters. It is straightforward to check that if we insert the  ansatz shown in Eq. (\ref{ansatz}) into Eq. (\ref{bianchi-I}) then we will obtain the following power-law scale factors. Note again that $\sigma$ stands for a deviation from isotropy and therefore should be much smaller than the isotropic scale factor $\alpha$, i.e., $\alpha \gg \sigma$ or equivalently $\zeta \gg \eta$ due to Eq. (\ref{ansatz}). As a result, the  field equations (\ref{Friedmann-equation}), (\ref{alpha-equation}), (\ref{sigma-equation}), (\ref{scalar-field-equation-1}), and (\ref{scalar-field-equation-2}), which are differential equations in time, can become a set of algebraic equations:
\begin{align}\label{algebraic-1}
\zeta^2 &= \eta^2 +\frac{1}{3}\left(\frac{\xi_\phi^2}{2}-\frac{\xi_\psi^2}{2}+\frac{\omega_0}{2} \xi_\phi \xi_\psi + u_\phi +u_\psi+ \frac{v}{2} \right), \\
\label{algebraic-2}
-\zeta &= -3\zeta^2 +u_\phi +u_\psi+ \frac{v}{6}, \\
\label{algebraic-3}
-\eta &= -3\zeta \eta +\frac{v}{3},\\
\label{algebraic-4}
-\xi_\phi - \frac{\omega_0}{2} \xi_\psi &= -3\zeta \left(\xi_\phi +\frac{\omega_0}{2} \xi_\psi \right) - \lambda_\phi u_\phi +\rho_\phi v, \\
\label{algebraic-5}
-\xi_\psi + \frac{\omega_0}{2} \xi_\phi &= -3\zeta \left(\xi_\psi -\frac{\omega_0}{2} \xi_\phi \right) + \lambda_\psi u_\psi - \rho_\psi v.
\end{align}
It is noted that in order to derive the above algebraic equations we have used the following constraints for the field parameters:
\begin{align}\label{constraint-1}
 \lambda_\phi \xi_\phi&=-2,
\\ \label{constraint-2}
\lambda_\psi \xi_\psi&=-2,
\\ \label{constraint-3}
\rho_\phi \xi_\phi+\rho_\psi \xi_\psi+2\zeta+2\eta &=1,
\end{align}
which make all terms in the field equations proportional to $t^{-2}$.
Note also that  we have introduced additional positive variables:
\begin{align}
 u_\phi &= \frac{{V_{0\phi} }}
{{M_p^2 }}\exp\left[{\lambda_\phi \phi _0 }\right] >0 , \\
u_\psi &= \frac{{V_{0\psi} }}
{{M_p^2 }}\exp\left[{\lambda_\psi \psi _0 }\right] >0, \\ 
v & = \frac{{p_A^2 f_0^{ - 2}  }}
{{M_p^2 }}\exp\left[{ - 2(\rho_\phi \phi _0+\rho_\psi \psi_0) }\right] >0,
\end{align}
for convenience. For the slow-roll inflation, which is based on the slow-roll approximation mentioned above, it turns out that
\begin{equation} \label{slow-roll}
u_\phi \gg \frac{2}{\lambda_\phi^2}, ~ u_\psi \gg \frac{2}{\lambda_\psi^2}, ~ u_\phi +u_\psi \gg v/2.
\end{equation}
 It is apparent that 
\begin{equation} \label{relation-1}
v = 3 \left(3\zeta-1\right)\eta,
\end{equation}
by noting Eq. (\ref{algebraic-3}). Hence, the positivity of $v$ leads to a constraint that $\eta >0$ during the inflationary phase with $\zeta \gg 1$. From the constraint equation (\ref{constraint-3}), we obtain that
\begin{equation}\label{relation-2}
\eta = -\zeta + \frac{\rho_\phi}{\lambda_\phi}+\frac{\rho_\psi}{\lambda_\psi} +\frac{1}{2}.
\end{equation}
Hence, the constraint that $\eta >0$ implies that
\begin{equation}
\zeta < \frac{\rho_\phi}{\lambda_\phi}+\frac{\rho_\psi}{\lambda_\psi} +\frac{1}{2}.
\end{equation}
It it noted that $\zeta \gg 1$ for the inflationary solutions. This will only be satisfied if 
\begin{equation}
\rho_\phi \gg \lambda_\phi, ~\rho_\psi \gg \lambda_\psi,
\end{equation}
assuming that all these $\lambda_{\phi,\psi}$ and $\rho_{\phi,\psi}$ are positive parameters. Thanks to the relations shown in Eqs. (\ref{relation-1}) and (\ref{relation-2}), the other equations (\ref{algebraic-4}) and (\ref{algebraic-5}) can be further reduced to
\begin{align}
&2 \lambda _{\psi } \lambda _{\phi }^2 u_\phi + \left(3 \zeta -1 \right) \left(6 \zeta  \lambda _{\psi } \lambda _{\phi } \rho _{\phi }-6 \lambda _{\psi } \rho _{\phi }^2-3 \lambda _{\psi } \lambda _{\phi } \rho _{\phi }-6 \lambda _{\phi } \rho _{\psi } \rho _{\phi }-4 \lambda _{\psi }-2 \omega _0 \lambda _{\phi }\right)=0,\\
&2 \lambda _{\psi }^2 \lambda _{\phi } u_\psi + \left(3 \zeta -1\right) \left(6 \zeta  \lambda _{\psi } \lambda _{\phi } \rho _{\psi }-6 \lambda _{\phi } \rho _{\psi }^2-3 \lambda _{\psi } \lambda _{\phi } \rho _{\psi }-6 \lambda _{\psi } \rho _{\psi } \rho _{\phi }+4 \lambda _{\phi }-2 \omega _0 \lambda _{\psi }\right)=0,
\end{align}
respectively, which can be solved to give non-trivial solutions of $u_\phi$ and $u_\psi$:
\begin{align} \label{u1}
u_\phi &=-\frac{1}{2 \lambda _{\psi } \lambda _{\phi }^2} \left(3 \zeta -1 \right) \left(6 \zeta  \lambda _{\psi } \lambda _{\phi } \rho _{\phi }-6 \lambda _{\psi } \rho _{\phi }^2-3 \lambda _{\psi } \lambda _{\phi } \rho _{\phi }-6 \lambda _{\phi } \rho _{\psi } \rho _{\phi }-4 \lambda _{\psi }-2 \omega _0 \lambda _{\phi }\right), \\
\label{u2}
u_\psi &= -\frac{1}{2 \lambda _{\psi }^2 \lambda _{\phi }}\left(3 \zeta -1\right) \left(6 \zeta  \lambda _{\psi } \lambda _{\phi } \rho _{\psi }-6 \lambda _{\phi } \rho _{\psi }^2-3 \lambda _{\psi } \lambda _{\phi } \rho _{\psi }-6 \lambda _{\psi } \rho _{\psi } \rho _{\phi }+4 \lambda _{\phi }-2 \omega _0 \lambda _{\psi }\right).
\end{align}
Given the solutions shown in Eqs. (\ref{relation-1}), (\ref{relation-2}), (\ref{u1}), and (\ref{u2}), we can obtain the following non-trivial equation of $\zeta$ from either Eq. (\ref{algebraic-1}) or Eq. (\ref{algebraic-2}):
\begin{align}
&-6\lambda_\phi \lambda_\psi \left( \lambda_\phi \lambda_\psi +2 \lambda _{\phi } \rho _{\psi }+2 \lambda _{\psi }  \rho _{\phi } \right) \zeta \nonumber\\
&+4 \left(\lambda_\phi \rho_\psi +\lambda_\psi \rho_\phi\right) \left(2\lambda_\phi \lambda_\psi +3\lambda_\phi \rho_\psi +3\lambda_\psi \rho_\phi\right) +\lambda_\phi^2 \lambda_\psi^2 +8 \left(\lambda_\psi^2 +\omega_0 \lambda_\phi \lambda_\psi -\lambda_\phi^2\right) =0,
\end{align}
which admits a non-trivial solution of $\zeta$:
\begin{equation}
\zeta = \frac{4 \left(\lambda_\phi \rho_\psi +\lambda_\psi \rho_\phi\right) \left(2\lambda_\phi \lambda_\psi +3\lambda_\phi \rho_\psi +3\lambda_\psi \rho_\phi\right) +\lambda_\phi^2 \lambda_\psi^2 +8 \left(\lambda_\psi^2 +\omega_0 \lambda_\phi \lambda_\psi -\lambda_\phi^2\right) }{6\lambda_\phi \lambda_\psi \left( \lambda_\phi \lambda_\psi +2 \lambda _{\phi } \rho _{\psi }+2 \lambda _{\psi }  \rho _{\phi } \right)}.
\end{equation}
Thanks to this explicit expression defined in terms of the field parameter $\lambda_{\phi,\psi}$ and $\rho_{\phi,\psi}$, the other field variables $\eta$, $u_\phi$, $u_\psi$, and $v$ now become as follows
\begin{align}
\eta =~&\frac{\lambda_\phi \lambda_\psi \left(\lambda_\phi \lambda_\psi +2\lambda_\phi \rho_\psi +2\lambda_\psi \rho_\phi \right) - 4\left(\lambda_\psi^2 +\omega_0 \lambda_\phi \lambda_\psi -\lambda_\phi^2\right)}{3 \lambda _{\phi } \lambda_\psi \left(\lambda _{\phi }\lambda _{\psi } +2 \lambda _{\phi } \rho _{\psi }+2 \lambda _{\psi } \rho _{\phi }\right)}, \\
u_\phi = ~&\Omega_0  \left[\lambda_\psi^2 \left(\lambda_\phi \rho_\phi +2\rho_\phi^2+2 \right)+2\lambda_\phi \lambda_\psi \rho_\phi \rho_\psi +4 \left(\lambda_\phi \rho_\phi +\lambda_\psi \rho_\psi\right) \right.\nonumber\\
&\left.+\omega_0 \left(\lambda_\phi \lambda_\psi +2\lambda_\phi \rho_\psi -2\lambda_\psi \rho_\phi\right)\right] ,\\
u_\psi =~&\Omega_0 \left[\lambda_\phi^2 \left(\lambda_\psi \rho_\psi +2\rho_\psi^2-2 \right)+2\lambda_\phi \lambda_\psi \rho_\phi \rho_\psi -4 \left(\lambda_\phi \rho_\phi +\lambda_\psi \rho_\psi\right)  \right.\nonumber\\
&\left. +\omega_0 \left(\lambda_\phi \lambda_\psi -2\lambda_\phi \rho_\psi +2\lambda_\psi \rho_\phi\right)\right] ,\\
v =~&\Omega_0 \left[\lambda_\phi \lambda_\psi \left(\lambda_\phi \lambda_\psi +2\lambda_\phi \rho_\psi +2\lambda_\psi \rho_\phi\right) -4\left(\lambda_\psi^2 +\omega_0 \lambda_\phi \lambda_\psi -\lambda_\phi^2\right)\right] ,
\end{align}
with an additional variable $\Omega_0$, whose value is given by
\begin{equation}
\Omega_0 = \frac{4 \left(\lambda_\phi \rho_\psi +\lambda_\psi \rho_\phi\right) \left(\lambda_\phi \lambda_\psi +3\lambda_\phi \rho_\psi +3\lambda_\psi \rho_\phi\right) -\lambda_\phi^2 \lambda_\psi^2 +8 \left(\lambda_\psi^2 +\omega_0 \lambda_\phi \lambda_\psi -\lambda_\phi^2\right)}{2 \left[\lambda _{\phi } \lambda_\psi \left(\lambda _{\phi }\lambda _{\psi } +2 \lambda _{\phi } \rho _{\psi }+2 \lambda _{\psi } \rho _{\phi }\right)\right]^2}.
\end{equation}
As discussed in Ref. \refcite{WFK}, these variables can be approximated as
\begin{align} \label{appro-1}
\zeta & \simeq  \frac{\rho_\phi}{\lambda_\phi} +\frac{\rho_\psi}{\lambda_\psi} \gg 1 ,\\
\label{appro-2}
\eta & \simeq \frac{1}{3},\\
\label{appro-3}
u_\phi & \simeq 3\frac{\rho_\phi}{\lambda_\phi} \left(\frac{\rho_\phi}{\lambda_\phi} +\frac{\rho_\psi}{\lambda_\psi}\right) \simeq 3\frac{\rho_\phi}{\lambda_\phi} \zeta \gg 1 ,\\
\label{appro-4}
u_\psi & \simeq 3\frac{\rho_\psi}{\lambda_\psi} \left(\frac{\rho_\phi}{\lambda_\phi} +\frac{\rho_\psi}{\lambda_\psi}\right) \simeq 3\frac{\rho_\psi}{\lambda_\psi} \zeta \gg 1,\\
\label{appro-5}
v & \simeq 3 \left(\frac{\rho_\phi}{\lambda_\phi} +\frac{\rho_\psi}{\lambda_\psi}\right) \simeq 3\zeta \gg 1,
\end{align}
during the inflationary phase, in which 
\begin{align}
\rho_\phi \gg \lambda_\phi \sim {\cal O}(1); ~\rho_\psi \gg \lambda_\psi \sim {\cal O}(1).
\end{align}
It is straightforward to see that these values satisfy the slow-roll approximation shown in Eq. (\ref{slow-roll}). Hence, the following slow-roll parameter, $\epsilon \equiv -\dot H/H^2$, can be defined to be
\begin{align}
\epsilon & = \frac{6\lambda_\phi \lambda_\psi \left( \lambda_\phi \lambda_\psi +2 \lambda _{\phi } \rho _{\psi }+2 \lambda _{\psi }  \rho _{\phi } \right)}{4 \left(\lambda_\phi \rho_\psi +\lambda_\psi \rho_\phi\right) \left(2\lambda_\phi \lambda_\psi +3\lambda_\phi \rho_\psi +3\lambda_\psi \rho_\phi\right) +\lambda_\phi^2 \lambda_\psi^2 +8 \left(\lambda_\psi^2 +\omega_0 \lambda_\phi \lambda_\psi -\lambda_\phi^2\right) }\nonumber\\
& \simeq \frac{\lambda_\phi \lambda_\psi}{\lambda_\phi \rho_\psi + \lambda_\psi \rho_\phi} \ll 1.
\end{align}
along with the value of the anisotropy parameter $\Sigma/H \equiv \dot\sigma/\dot\alpha$:
\begin{align}
\frac{\Sigma}{H} & = \frac{2\left[\lambda_\phi \lambda_\psi \left(\lambda_\phi \lambda_\psi +2\lambda_\phi \rho_\psi +2\lambda_\psi \rho_\phi \right) - 4\left(\lambda_\psi^2 +\omega_0 \lambda_\phi \lambda_\psi -\lambda_\phi^2\right) \right]}{4 \left(\lambda_\phi \rho_\psi +\lambda_\psi \rho_\phi\right) \left(2\lambda_\phi \lambda_\psi +3\lambda_\phi \rho_\psi +3\lambda_\psi \rho_\phi\right) +\lambda_\phi^2 \lambda_\psi^2 +8 \left(\lambda_\psi^2 +\omega_0 \lambda_\phi \lambda_\psi -\lambda_\phi^2\right) } \nonumber\\
& \simeq \frac{\lambda_\phi \lambda_\psi}{3\left(\lambda_\phi \rho_\psi + \lambda_\psi \rho_\phi \right)} \ll 1.
\end{align}
It is clear that the anisotropy parameter is really small for the Bianchi type I inflationary solutions. This result turns out to be consistent with that investigated in the previous works, where the mixed term of canonical and phantom scalar fields has not been shown up \cite{WFK}.  This result is also consistent with the recent observational data of the WMAP \cite{WMAP} and Planck  \cite{Planck}. In other words, the mixed term plays a little role  in the field equations during the slow-roll inflationary phase so that it does not affect much on the obtained solutions as well as the anisotropy parameter. Below, we will see if the mixed term could change the stability of the Bianchi type I solutions during the inflationary phase. Once again, we note that  all above power-law solutions will reduce to that found in the two-scalar-field model in Ref. \refcite{WFK} once the limit $\omega_0 \to 0$ is taken. 
\section{Stability analysis of the inflationary Bianchi type I solutions} \label{sec3}
Given the above anisotropic power-law solutions, we now would like to examine their stability during the inflationary phase with $\zeta \gg 1$ in order to test the validity of the cosmic no-hair conjecture proposed by Hawking and his colleagues \cite{Hawking}, which has been proved partially for the Bianchi spaces by Wald \cite{wald}. It is noted that the Wald's proof deals only with the dominant and strong energy conditions and therefore it could not provide us full information about the stability of inflationary solutions, which could only be improved by perturbation analysis, e.g., the dynamical system \cite{KSW,WFK,extensions} or  power-law perturbations \cite{WFK}.
 
Note that  the power-law perturbations have been used in Ref. \refcite{WFK} in order to examine the validity of the cosmic no-hair conjecture since they are compatible with the following anisotropic power-law solutions.  However, we have also pointed out in Ref. \refcite{WFK}  that the power-law perturbations approach is consistent with the dynamical system approach  for at least one-scalar-field models, e.g., the KSW model \cite{KSW} or its non-canonical extensions \cite{WFK}. In this paper, therefore, we would like to use the dynamical system method to investigate the stability of the anisotropic power-law solutions for the two-scalar-field model with the mixed term by introducing the following dynamical variables \cite{KSW,WFK}:
\begin{align} \label{dynamical-variables}
X&=\frac{\dot\sigma}{\dot\alpha},~ Y_\phi=\frac{\dot\phi}{M_p \dot\alpha},~ Y_\psi=\frac{\dot\psi}{M_p \dot\alpha},\nonumber\\
Z&=\frac{f^{-1}(\phi,\psi)}{M_p \dot\alpha}\exp[-2\alpha-2\sigma]p_A, \nonumber\\
W_\phi &= \frac{\sqrt{V_\phi}}{M_p\dot\alpha}, ~W_\psi =\frac{\sqrt{V_\psi}}{M_p\dot\alpha},
\end{align}
here $W_\phi$ and $W_\psi$ are auxiliary dynamical variables, which will be useful for further calculations on autonomous equations \cite{quintom1}.
As a result, the derivatives of these dynamical variables with respect to $\alpha$ acting as a new time coordinate, $d\alpha =\dot\alpha dt$ \cite{KSW,WFK}, turn out to be
\begin{align}
 \frac{dX}{d\alpha}&=\frac{\ddot\sigma}{\dot\alpha^2}-\frac{\ddot\alpha}{\dot\alpha^2}X,
\\ 
\frac{d Y_\phi}{d\alpha}&=\frac{\ddot\phi}{M_p \dot\alpha^2}-\frac{\ddot\alpha}{\dot\alpha^2}Y_\phi,
\\ 
\frac{d Y_\psi}{d\alpha}&=\frac{\ddot\psi}{M_p \dot\alpha^2}-\frac{\ddot\alpha}{\dot\alpha^2}Y_\psi,
\\
\frac{dZ}{d\alpha}&=-\left(\rho_\phi Y_\phi + \rho_\psi Y_\psi \right)Z- 2\left({X+1}\right)Z-\frac{\ddot\alpha}{\dot\alpha^2}Z, \\
\frac{dW_\phi}{d\alpha}&= \left(\frac{\lambda_\phi}{2}Y_\phi-\frac{\ddot\alpha}{\dot\alpha^2}\right)W_\phi,\\
\frac{dW_\psi}{d\alpha}&= \left(\frac{\lambda_\psi}{2}Y_\phi-\frac{\ddot\alpha}{\dot\alpha^2}\right)W_\psi,
\end{align}
here the exponential forms of $V_\phi$, $V_\psi$, and $f(\phi,\psi)$ shown in the previous section have been used.
Thanks to these explicit expressions, the field equations (\ref{Friedmann-equation}), (\ref{alpha-equation}), (\ref{sigma-equation}), (\ref{scalar-field-equation-1}), and (\ref{scalar-field-equation-2}) can be transformed into the following autonomous equations:
\begin{align} \label{autonomous-1}
\frac{dX}{d\alpha} =& \left[3\left(X^2-1\right) +\frac{1}{2} \left(Y_\phi^2 +\omega_0 Y_\phi Y_\psi - Y_\psi^2\right) +\frac{Z^2}{3}\right] X +\frac{Z^2}{3}  ,\\
\label{autonomous-2}
\frac{d Y_\phi}{d\alpha} +\frac{\omega_0}{2}\frac{d Y_\psi}{d\alpha} =& \left[3\left(X^2-1\right) +\frac{1}{2}\left(Y_\phi^2 +\omega_0 Y_\phi Y_\psi - Y_\psi^2\right) +\frac{Z^2}{3}\right] \left(Y_\phi +\frac{\omega_0}{2}Y_\psi\right) \nonumber\\
&-\lambda_\phi W_\phi^2 +\rho_\phi Z^2, \\
\label{autonomous-3}
\frac{d Y_\psi}{d\alpha} -\frac{\omega_0}{2}\frac{d Y_\phi}{d\alpha} =& \left[3\left(X^2-1\right) +\frac{1}{2}\left(Y_\phi^2 +\omega_0 Y_\phi Y_\psi - Y_\psi^2\right) +\frac{Z^2}{3}\right] \left(Y_\psi -\frac{\omega_0}{2}Y_\phi\right) \nonumber\\
&+\lambda_\psi W_\psi^2 -\rho_\psi Z^2,\\
\label{autonomous-4}
\frac{dZ}{d\alpha}=&~Z \Bigl[3\left(X^2-1\right) +\frac{1}{2}\left(Y_\phi^2 +\omega_0 Y_\phi Y_\psi - Y_\psi^2\right) +\frac{Z^2}{3} -2 X \nonumber\\
&  - \left(\rho_\phi Y_\phi +\rho_\psi Y_\psi \right) +1 \Bigr],
\end{align}
\begin{align}
\label{autonomous-5}
\frac{d W_\phi}{d\alpha}=&\left[3X^2  +\frac{1}{2}\left(Y_\phi^2 +\omega_0 Y_\phi Y_\psi - Y_\psi^2\right) +\frac{Z^2}{3} +\frac{\lambda_\phi}{2}Y_\phi\right] W_\phi, \\
\label{autonomous-6}
\frac{d W_\psi}{d\alpha}=&\left[3X^2  +\frac{1}{2}\left(Y_\phi^2 +\omega_0 Y_\phi Y_\psi - Y_\psi^2\right) +\frac{Z^2}{3} +\frac{\lambda_\psi}{2}Y_\psi\right] W_\psi,
\end{align}
where we have used the result derived from the Friedmann equation (\ref{Friedmann-equation}) such as
\begin{equation} \label{dynamical-Friedmann}
W_\phi^2 +W_\psi^2= -3\left(X^2-1\right) -\frac{1}{2}\left(Y_\phi^2 +\omega_0 Y_\phi Y_\psi - Y_\psi^2 + Z^2 \right).
\end{equation}

Below, we will  show that anisotropic ($X\neq 0$) fixed point solutions of the autonomous equations are indeed equivalent to the obtained anisotropic power-law solutions of the Einstein field equations. Hence, the stability of anisotropic fixed points will tell us that of the corresponding anisotropic power-law solutions. It is known that anisotropic fixed points of the dynamical system are solutions of the following equations: 
\begin{equation}\label{fixed-point-equations}
\frac{d X}{d\alpha} =\frac{d Y_\phi}{d\alpha} = \frac{d Y_\psi} { d\alpha} =\frac{dZ}{d\alpha} = \frac{dW_\phi }{d\alpha} =\frac{dW_\psi}{d\alpha} =0.
\end{equation}
 As a result, from the last two equations in Eq. (\ref{fixed-point-equations}), i.e., $dW_\phi /d\alpha =dW_\psi/d\alpha =0$, we obtain the following relations for the dynamical variables:
\begin{equation} \label{relation-Yphi-Ypsi}
3X^2  +\frac{1}{2}\left(Y_\phi^2 +\omega_0 Y_\phi Y_\psi - Y_\psi^2\right) +\frac{Z^2}{3} =-\frac{\lambda_\phi}{2}Y_\phi =-\frac{\lambda_\psi}{2}Y_\psi,
\end{equation}
requiring that $W_\phi \neq 0$, $W_\psi \neq 0$, and $Z\neq0$ for non-trivial fixed points. Furthermore, using these relations will lead the other equations for the fixed points to
\begin{align}
\label{fixed-point-eq-1}
\left(\frac{\lambda_\phi}{2} Y_\phi +3 \right) X -\frac{Z^2}{3}&=0,\\
\label{fixed-point-eq-2}
-\left(\frac{\lambda_\phi}{2} Y_\phi +3\right) \left(1 +\frac{\omega_0}{2}\frac{\lambda_\phi}{\lambda_\psi}\right) Y_\phi-\lambda_\phi W_\phi^2 +\rho_\phi Z^2&=0, \\
\label{fixed-point-eq-3}
-\left(\frac{\lambda_\phi}{2} Y_\phi +3\right)\left(\frac{\lambda_\phi}{\lambda_\psi} -\frac{\omega_0}{2}\right) Y_\phi+\lambda_\psi W_\psi^2 -\rho_\psi Z^2&=0,\\
\label{fixed-point-eq-4}
2X +\left(\frac{\lambda_\phi}{2} +\rho_\phi +\frac{\lambda_\phi}{\lambda_\psi} \rho_\psi\right) Y_\phi +2 &=0.
\end{align}
Next, we will eliminate the existence of $W_\phi$ and $W_\psi$ in Eqs. (\ref{fixed-point-eq-2}) and (\ref{fixed-point-eq-3}) with the help of Eq. (\ref{dynamical-Friedmann}) to obtain the following equation:
\begin{equation}\label{fixed-point-eq-5}
-\left(\frac{\lambda_\phi Y_\phi}{2}+3\right) \left[ \left(\lambda_\psi  -\frac{\lambda_\phi^2 }{\lambda_\psi}+\omega_0 \lambda_\phi  \right)Y_\phi  +\lambda_\phi \lambda_\psi \right] +\left(\frac{\lambda_\phi \lambda_\psi }{6}+\lambda_\psi \rho_\phi+\lambda_\phi \rho_\psi  \right)Z^2 =0.
\end{equation}
It is apparent that we now have three independent equations (\ref{fixed-point-eq-1}), (\ref{fixed-point-eq-4}), and (\ref{fixed-point-eq-5}) for three dynamical variables $X$, $Y_\phi$, and $Z^2$. Indeed, solving these non-linear equations gives us two solutions, one is trivial corresponding to $Z=0$ and the other is a non-trivial fixed point given by
\begin{align}
X&=\frac{2}{Q}\left[\lambda_\phi \lambda_\psi \left(\lambda_\phi \lambda_\psi +2 \lambda_\phi \rho_\psi +2 \lambda_\psi \rho_\phi \right) -4\left(\lambda_\psi^2 +\omega_0 \lambda_\phi \lambda_\psi -\lambda_\phi^2 \right) \right],\\
Y_\phi&=-\frac{12}{Q} \lambda _{\psi } \left(\lambda_\phi \lambda_\psi +2 \lambda_\phi \rho_\psi +2 \lambda_\psi \rho_\phi \right),\\
Z^2 &=\frac{18 }{Q^2} \hat \Omega_0 \left[\lambda_\phi \lambda_\psi \left(\lambda_\phi \lambda_\psi+2\lambda_\phi \rho_\psi +2\lambda_\psi \rho_\phi\right)-4 \left(\lambda_\psi^2 +\omega_0 \lambda_\phi \lambda_\psi -\lambda_\phi^2\right)\right] ,
\end{align}
with
\begin{align}
&\hat\Omega_0 \equiv 4\left(\lambda_\phi \rho_\psi +\lambda_\psi \rho_\phi\right)\left(\lambda_\phi \lambda_\psi +3\lambda_\phi \rho_\psi +3\lambda_\psi \rho_\phi \right) -\lambda_\phi^2 \lambda_\psi^2 +8\left(\lambda_\psi^2 +\omega_0 \lambda_\phi \lambda_\psi -\lambda_\phi^2 \right),\\
&Q\equiv 4\left(\lambda_\phi \rho_\psi +\lambda_\psi \rho_\phi\right) \left(2\lambda_\phi \lambda_\psi + 3\lambda_\phi \rho_\psi +3\lambda_\psi \rho_\phi \right) +\lambda_\phi^2 \lambda_\psi^2 + 8 \left(\lambda_\psi^2 +\omega_0 \lambda_\phi \lambda_\psi -\lambda_\phi^2 \right).
\end{align}
 Note that $Y_\psi$ can be defined in terms of $Y_\phi$ as shown in Eq. (\ref{relation-Yphi-Ypsi}). It is straightforward to see that these anisotropic fixed point solutions are indeed equivalent to the anisotropic power-law solutions found in the previous section. Indeed, one can easily obtain the above expressions of the dynamical variables by using the following relations:
\begin{equation}
X=\frac{\eta}{\zeta}, ~ Y_\phi = -\frac{2}{\lambda_\phi \zeta}, ~Y_\psi = -\frac{2}{\lambda_\psi \zeta}, ~ Z^2 =\frac{v}{\zeta^2}, ~ W_\phi^2 =\frac{u_\phi}{\zeta^2}, ~ W_\psi^2 =\frac{u_\psi}{\zeta^2},
\end{equation}
with the explicit values $\zeta$, $\eta$, and $v$ in terms of $\lambda_{\phi,\psi}$ and $\rho_{\phi,\psi}$ have been derived in the previous section for the anisotropic power-law solutions.
Hence, the stability of the anisotropic power-law solutions can be determined by considering that of the corresponding anisotropic fixed points.  

As a result, during the inflationary phase we have  $\zeta \simeq  \lambda_\phi /\rho_\phi + \lambda_\psi /\rho_\psi \gg 1$, $\eta \simeq 1/3$, $u_\phi \simeq 3 (\rho_\phi /\lambda_\phi) \zeta \gg 1$, $u_\psi \simeq 3 (\rho_\psi /\lambda_\psi) \zeta \gg 1$, and $v\simeq 3\zeta \gg 1$, assuming that $\rho_\phi \gg \lambda_\phi \sim {\cal O}(1)$ along with  $\rho_\psi \gg \lambda_\psi \sim {\cal O}(1)$. Therefore, the inflationary anisotropic fixed points behave as follows $X, ~Y_\phi,~Y_\psi, ~Z^2 \ll 1$ and $W_\phi^2 \sim W_\psi^2 \sim 3/2$ assuming that $\rho_\phi /\lambda_\phi \sim \rho_\psi /\lambda_\psi$. 

Given the above results, we now would like to perturb the autonomous equations (\ref{autonomous-1})-(\ref{autonomous-6}) around the obtained anisotropic fixed points during the inflationary phase. As a result, the following perturbation  equations are approximately defined to be
\begin{align}
\frac{d \delta X}{d \alpha} & \simeq -3\delta X, \\
\frac{d \delta Y_\phi}{d\alpha} +\frac{\omega_0}{2}\frac{d\delta Y_\psi}{d\alpha} & \simeq -3 \left(\delta Y_\phi +\frac{\omega_0}{2}\delta Y_\psi \right) - 2\lambda_\phi W_\phi \delta W_\phi + \rho_\phi \delta Z, \\
\frac{d \delta Y_\psi}{d\alpha} - \frac{\omega_0}{2}\frac{d\delta Y_\phi}{d\alpha} & \simeq -3 \left(\delta Y_\psi - \frac{\omega_0}{2}\delta Y_\phi \right) + 2\lambda_\psi W_\psi \delta W_\psi - \rho_\psi \delta Z, \\
\frac{d \delta Z}{d\alpha} & \simeq - Z \left(2\delta X + \rho_\phi \delta Y_\phi + \rho_\psi Y_\psi \right), \\
\frac{d\delta W_\phi}{d\alpha} &\simeq \frac{\lambda_\phi}{2}W_\phi \delta Y_\phi ,\\
\frac{d\delta W_\psi}{d\alpha} &\simeq \frac{\lambda_\psi}{2} W_\psi \delta Y_\psi,
\end{align}
here we have only kept the leading terms in the above perturbation equations for simplicity. 
Now, taking the exponential perturbations for the dynamical variables as \cite{KSW,WFK}
\begin{align}
&\delta X = A_X \exp\left[\omega \alpha\right], ~ \delta Y_\phi = A_{Y_\phi} \exp\left[\omega \alpha\right], ~ \delta Y_\psi = A_{Y_\psi} \exp\left[\omega \alpha\right],\\
&\delta Z = A_Z \exp\left[\omega \alpha\right], ~ \delta W_\phi = A_{W_\phi} \exp\left[\omega \alpha\right], ~ \delta W_\psi = A_{W_\psi} \exp\left[\omega \alpha\right],
\end{align}
will lead the above perturbation equations to the following algebraic equations, which can be written as a matrix equation:
\begin{equation} \label{stability-equation}
{\cal H}\left( {\begin{array}{*{20}c}
   A_X  \\
   A_{Y_\phi}  \\
   A_{Y_\psi} \\
   A_Z  \\
   A_{W_\phi}\\
   A_{W_\psi}\\

 \end{array} } \right) \equiv \left[ {\begin{array}{*{20}c}
   {-\omega-3} & {0} & {0 } & {0 } & {0} &{0}  \\
   {0 } & {-3-\omega} & {-\frac{\omega_0}{2} \left(3+\omega\right) } & {\rho_\phi } &{-2\lambda_\phi W_\phi}&{0}  \\
     {0 } & {\frac{\omega_0}{2}\left(3+\omega\right)} & {-3-\omega } & {-\rho_\psi } &{0}&{2\lambda_\psi W_\psi}  \\
   {-2Z} & {-\rho_\phi Z } & {-\rho_\psi Z } & {-\omega } &{0}&{0} \\
   {0}&{\frac{\lambda_\phi W_\phi}{2}}&{0} &{0} &{-\omega}&{0}\\
   {0}&{0}&{\frac{\lambda_\psi W_\psi}{2}} &{0} &{0}&{-\omega}\\

 \end{array} } \right]\left( {\begin{array}{*{20}c}
   A_X  \\
   A_{Y_\phi}  \\
   A_{Y_\psi} \\
   A_Z  \\
   A_{W_\phi}\\
   A_{W_\psi}\\

 \end{array} } \right) = 0.
\end{equation}
Mathematically, the equation (\ref{stability-equation}) admits non-trivial solutions if and only if
\begin{equation}
\det {\cal H}=0,
\end{equation}
which can be evaluated to be a polynomial equation of $\omega$ as
\begin{equation} \label{poly}
\omega f(\omega)\equiv \omega \left( a_6 \omega^5 + ...+a_1 \right) =0,
\end{equation}
where 
\begin{align}
a_6 &= \frac{\omega _0^2}{4}+1 >0,\\
\label{def.of.a1}
a_1 &= -3 \left(\lambda _{\psi }^2 \lambda _{\phi }^2 W_{\psi }^2 W_{\phi }^2+ \lambda _{\psi }^2 \rho _{\phi }^2 W_{\psi }^2 Z+ \lambda _{\phi }^2 \rho _{\psi }^2  W_{\phi }^2Z\right) <0.
\end{align}
Here, we do not show the expressions of $a_i$'s ~($i=2-5$) because we only want to know the sign of the highest power term, $a_6$, and that of the lowest power term, $a_1$. The reason is based on the observation mentioned in Ref. \refcite{WFK}. In particular, we have observed that if $a_6 >0$ and $a_1<0$ (or inversely $a_6 <0$ and $a_1>0$) then the polynomial equation of $\omega$, $f(\omega)=0$ as shown in Eq. (\ref{poly}), will admit at least one positive root $\omega>0$, which corresponds to a unstable perturbation mode for the anisotropic fixed points. More specifically, this claim can be understood as follows: $f(\omega) \sim a_6 \omega^5 >0 $ as $\omega \gg 1$ and $f(\omega=0)=a_1<0$, then the curve $f(\omega)$ will cross the positive horizontal $\omega$-axis at least one time at $\omega=\omega^\ast$, and this intersection point $\omega=\omega^\ast$ is indeed a positive root to the equation  $f(\omega)=0$. 

It appears that the coefficient $a_6$ is always positive whatever the sign of $\omega_0$. In addition, the sign of $a_1$ is independent of that of $\omega_0$. Hence, we can conclude that the inclusion of extra mixed term, $\omega_0 \partial_\mu\phi \partial^\mu \psi$, does not change the stability of the two-scalar-field model \cite{WFK}. 
\section{Absence of pure kinetic terms of scalar fields} \label{sec4}
In this section, we would like to discuss a special scenario, in which the pure kinetic terms of scalar fields will not show up, i.e., $a=b=0$, leaving only the mixed kinetic term of scalar fields in the following action:
\begin{align}
  S = \int {d^4 } x\sqrt { - g} & \left[ {\frac{M_p^2}
{2}R -\frac{\omega_0}{2} \partial^\mu \phi \partial_\mu \psi - V_\phi(\phi) -V_\psi(\psi) - \frac{1}{4}f^2( \phi,\psi )F_{\mu \nu } F^{\mu \nu } } \right].
\end{align}
 In this scenario, it turns out that we cannot distinguish $\phi$ and $\psi$ as canonical or phantom fields although their mixed kinetic term can be either canonical or phantom depending on the sign of $\omega_0$. In particular, $\omega_0>0$ and $\omega_0<0$ will correspond to the quintessence-like and phantom-like mixed terms, respectively. As a result, the corresponding  anisotropic power-law solutions will be solved to be
\begin{align}
\zeta =~& \frac{4 \left(\lambda_\phi \rho_\psi +\lambda_\psi \rho_\phi\right) \left(2\lambda_\phi \lambda_\psi +3\lambda_\phi \rho_\psi +3\lambda_\psi \rho_\phi\right) +\lambda_\phi^2 \lambda_\psi^2 +8\omega_0 \lambda_\phi \lambda_\psi}{6\lambda_\phi \lambda_\psi \left( \lambda_\phi \lambda_\psi +2 \lambda _{\phi } \rho _{\psi }+2 \lambda _{\psi }  \rho _{\phi } \right)},\\
\eta =~&\frac{\lambda_\phi \lambda_\psi +2\lambda_\phi \rho_\psi +2\lambda_\psi \rho_\phi - 4\omega_0}{3  \left(\lambda _{\phi }\lambda _{\psi } +2 \lambda _{\phi } \rho _{\psi }+2 \lambda _{\psi } \rho _{\phi }\right)}, \\
u_\phi = ~&\bar\Omega_0  \left[\lambda_\psi^2 \rho_\phi\left(\lambda_\phi +2\rho_\phi \right)+2\lambda_\phi \lambda_\psi \rho_\phi \rho_\psi +\omega_0 \left(\lambda_\phi \lambda_\psi +2\lambda_\phi \rho_\psi -2\lambda_\psi \rho_\phi\right)\right] ,\\
u_\psi =~&\bar\Omega_0 \left[\lambda_\phi^2 \rho_\psi \left(\lambda_\psi  +2\rho_\psi \right)+2\lambda_\phi \lambda_\psi \rho_\phi \rho_\psi +\omega_0 \left(\lambda_\phi \lambda_\psi -2\lambda_\phi \rho_\psi +2\lambda_\psi \rho_\phi\right)\right] ,\\
v =~&\bar\Omega_0 \lambda_\phi \lambda_\psi \left(\lambda_\phi \lambda_\psi +2\lambda_\phi \rho_\psi +2\lambda_\psi \rho_\phi -4\omega_0  \right) ,
\end{align}
with the value of $\bar\Omega_0$ is given by
\begin{equation}
\bar\Omega_0 = \frac{4 \left(\lambda_\phi \rho_\psi +\lambda_\psi \rho_\phi\right) \left(\lambda_\phi \lambda_\psi +3\lambda_\phi \rho_\psi +3\lambda_\psi \rho_\phi\right) -\lambda_\phi^2 \lambda_\psi^2 +8 \omega_0 \lambda_\phi \lambda_\psi }{2 \left[\lambda _{\phi } \lambda_\psi \left(\lambda _{\phi }\lambda _{\psi } +2 \lambda _{\phi } \rho _{\psi }+2 \lambda _{\psi } \rho _{\phi }\right)\right]^2}.
\end{equation}
Here, the initial setup of the fields, metric $g_{\mu\nu}$, and potentials has been remained.  
It turns out that we still obtain the anisotropic power-law solutions with non-vanishing $\eta$ in this case although the terms associated with the pure kinetic terms of $\phi$ and $\psi$ have not shown in the following Einstein field equations. It has been shown that these terms are very small compared to that involving $\rho_\phi$ and $\rho_\psi$  due to the requirement for the inflationary phase that $\rho_{\phi,\psi} \gg \lambda_{\phi,\psi} \sim {\cal O}(1)$. Hence, the  inflationary solutions can be approximated to be that shown in Eqs. (\ref{appro-1}), (\ref{appro-2}), (\ref{appro-3}), (\ref{appro-4}), and (\ref{appro-5}). 

Now, we would like to discuss the stability of the anisotropic solutions found in this section.  In particular, the stability analysis based on the dynamical system approach as shown in the previous section will be used again. As a result, we are able to obtain the new value for $a_6$ such as 
\begin{equation}
a_6=\frac{\omega_0^2}{4}>0,
\end{equation} 
while $a_1$ remains its negative value as shown in Eq. (\ref{def.of.a1}). This result clear indicates that the corresponding anisotropic solutions found in this special scenario are also unstable during the inflationary phase, no matter the sign of $\omega_0$. 
\section{Conclusions} \label{sec5}
It is noted that most of inflation models have worked for isotropic metrics such as the FLRW or de Sitter spacetime. However, some anomalies like the hemispherical asymmetry and the Cold Spot existing in the CMB temperature \cite{WMAP,Planck} address  necessary modifications to the inflationary models \cite{Buchert:2015wwr}. In particular, the anisotropic inflation dealing with anisotropic spacetimes, e.g., the Bianchi metrics, might be a better framework for investigating the properties of the early universe, where some exotic features might be explained \cite{Ade:2013vbw,Bianchi,Pitrou:2008gk}.  As mentioned above,  if the cosmic no-hair conjecture proposed by Hawking and his colleagues  holds, the state of universe at the late time should be isotropic, no matter the initial state of the universe~\cite{Hawking}.  Some people have tried to prove this conjecture for quite general metrics, e.g., the Bianchi spacetimes, which are homogeneous but anisotropic \cite{wald}, or the inhomogeneous and anisotropic ones \cite{inhomogeneous}. Otherwise, some people have tried to seek counter-examples to this conjecture by one way or the other \cite{KSW,KSW1,WFK,extensions}. In particular, the first correct counter-example to the conjecture has been found in the supergravity motivated model \cite{KSW}. As a result, this model has been proved to admit a stable and attractor Bianchi type I solutions during the inflationary phase, even when the scalar field is non-canonical~\cite{WFK}. In order to support the Hawking conjecture, some papers dealing with the inclusion of the phantom field, whose kinetic term is negative definite, have appeared~\cite{WFK}. However, we do not know whether this conjecture still survives in extended frameworks, where other unusual interaction terms of scalar fields, which might be allowed to appear in the early universe in order to explain the observational data \cite{mixed-1}, are considered. In this paper, therefore,  we would like to study the specific extended scenario of the two-scalar-field model \cite{WFK}, in which the mixed term of the canonical and phantom scalar fields \cite{mixed} is introduced. As a result, we have derived the following Bianchi type I power-law solutions for the studied model.  We have found that the anisotropy parameter is really small for inflationary solutions, consistent with the observational data of the WMAP \cite{WMAP} or Planck \cite{Planck}. The stability analysis has been performed to conclude that the obtained anisotropic solutions are indeed unstable during the inflationary phase, meaning that the mixed term does not affect, at least in the framework of two-scalar-field extension of KSW model, on the validity of the cosmic no-hair conjecture. Additionally, we have also considered the special case, in which the pure kinetic terms of $\phi$ and $\psi$ are ignored altogether in the field equations. As a result, the corresponding power-law solutions of this case also turn out to be unstable during the inflationary phase as expected.
 
These results along with that investigated in Ref. \refcite{WFK} strongly indicate that the unusual kinetic terms such as the kinetic term of phantom field and the mixed kinetic term of canonical and phantom fields  can play a significant role in protecting  the cosmic no-hair conjecture from the counter-examples admitted by the supergravity motivated model \cite{KSW}. In other words, the validity of cosmic no-hair conjecture seems to  require the existence of extra fields along with their interaction terms, which might exist in very high energy scales of the early universe. We would like to note that this paper has been devoted to investigate only the validity of the cosmic no-hair conjecture in the unusual scenarios associated with the inclusion of the mixed kinetic term of canonical and phantom fields. A detailed comparison with  the recent observations like the WMAP\cite{WMAP} or Planck\cite{Planck} for the two-scalar-field model studied in this paper  through correlations between $T$, $E$, and $B$ modes\cite{implication} should be done in further works and presented elsewhere. 

It is known that the loop quantum cosmology (gravity) has been regarded as one of leading theories to solve the singularity problems \cite{Singh:2011gp,singularity,Ashtekar:2011ni,Singh}, such as the Big Bang and Big Rip singularities, which have been very famous challenges to classical gravity, where quantum effects have been ignored. Hence, the cosmic no-hair conjecture should be tested not only in classical gravity framework but also in quantum gravity one, such as the loop quantum cosmology \cite{Ashtekar:2011ni,Singh}, in order to improve its cosmological validity. For example, the stability of Bianchi spacetimes should be investigated in the context of loop quantum cosmology \cite{Singh:2011gp} during the inflationary phase in order to see whether these spacetimes approach the de Sitter spacetime for large values of the time as predicted by the cosmic no-hair conjecture or not. 

We hope that our present paper would shed more light on the nature of early universe, especially the nature of observed anomalies  of the CMB temperature like the hemispherical asymmetry and the Cold Spot.
\section*{Acknowledgments}
T.Q.D. is deeply grateful to Professor W.~F.~Kao of Institute of Physics in National Chiao Tung University for his useful advice on the cosmic no-hair conjecture and the KSW anisotropic inflation model. We would like to thank an anonymous referee very much for useful comments. This research is supported in part by VNU University of Science, Vietnam National University, Hanoi. 

\end{document}